\documentclass[12pt]{article}
\usepackage[T1]{fontenc}
\usepackage{graphicx}

\newcommand{\ceq}[1]{(\ref{#1})}

\newfont{\mbld}{cmbx10 scaled 800}
\newfont{\cab}{cmsy10 scaled 1200}
\newfont{\scab}{cmsy10 scaled 1000}
\newfont{\bcall}{cmbsy10 scaled 1200}

\title{Cosmic ray recipes}
\author{Franco Ferrari\thanks{e-mail: ferrari@univ.szczecin.pl
  }$\enskip\,
$and
  Ewa Szuszkiewicz\thanks{e-mail: szusz@univ.szczecin.pl}
\\
\it Institute of Physics and CASA*, University of Szczecin,\\
\it   ul. Wielkopolska 15, 70-451 Szczecin, Poland.
}
\begin{document}
\maketitle
\begin{abstract}
Cosmic rays represent one of the most fascinating research
themes in modern
astronomy and physics.
After almost a century since their discovery, a huge amount of
scientific literature has been written on this topic and 
it is not always
easy to extract from it the necessary information
for somebody who approaches the subject for the first time.
This has been the main motivation  for preparing
 this article, which is a concise and self-contained review
 for whoever is interested in studying cosmic rays.
The priority has been given here to well established facts, which are
not at risk to get obsolete in a few years due to the fast progress of
the research in this field. Also many data are presented, which are
useful to characterize the doses of ionizing radiation 
delivered
to organisms living
on the Earth due to cosmic rays. 
The technical terms which are often encountered in
the scientific literature are explained in a separate appendix.
\end{abstract}

\section{Introduction}
Cosmic rays (CR) represent a fascinating subject of research, which is
recently witnessing a growing interest within the scientific
community.
With the next generation of cosmic ray detectors,
like
for example the Pierre Auger Observatory
which is currently under construction,
there is the hope
that the questions posed by
 these cosmic particles
will find soon some satisfactory answer. 
One should also note that the attention to
CR is not only restricted 
 to the traditional fields of high energy physics and astroparticles.
This article is for instance the result of the authors' efforts to
investigate the mutagenic effects of cosmic rays on cells. To this
purpose, one needs to characterize very precisely the fluxes and
intensities of particles arriving to the ground as an effect of the
interaction of CR with the atmosphere of the Earth.

The increasing popularity of CR is accompanied by an
increasing
 demand of information on this topic, but it is not always easy
to extract this information from the huge amount of literature which has been
written on CR during a period of almost a century after their
discovery.
An additional problem is that, looking at the scientific literature on
CR,
often concepts like the integral vertical intensity or the
differential integrated flux
are encountered.
These terms sound somewhat
puzzling and exotic for 
a reader that encounters them for the first time.
These motivations have compelled us to prepare this
 short review, which would like to be a compact but self-contained
reference on CR.
The preference has been given here
to well established facts, which are not
 in danger to become obsolete in a few years due to the 
 fast progress in this topic.
Moreover, a considerable effort has been done in order to explain in
 details 
 and with the 
 help of figures
 the technical terms used in the current 
scientific literature.
The puzzles raised by the existence of  ultra-high
energy cosmic rays with energy of 10$^{19}$eV or higher, like the
mystery of their origins or the
apparent violation of the theory of relativity which is connected with
them, 
have been just briefly mentioned.  
Hopefully, these puzzles will be solved with the next generation of
cosmic rays detectors.
Another aim  of this work
is to present data concerning the
physical parameters,
e.~g. types of
radiation, delivered effective doses and dose rates,
fluxes and intensities of incoming  particles, 
which
characterize the radiation to which the organisms living on
the Earth are exposed because of CR.
These data are certainly of interest for scientists working in life sciences.
Cosmic rays are in fact
the source of an almost uniform background of ionizing radiation
which is present everywhere on the Earth. 
Most of their energy arrives to the ground in the form of kinetic energy of
muons. The latter particles
are very penetrating and are able to
travel for 
kilometers in water and for hundreds of meters in rock. Since ionizing
radiation is mutagenic, it is very likely that the radiation of cosmic
origin has shaped in some way the evolution of life on our planet,
generating some kind of adaptive response in cells.

Finally, as already mentioned, this article is self-contained,
but of course it is far from being complete, because it is impossible
to cover all the immense literature which exists on cosmic rays.
To integrate the material presented here,
suggested further readings are for instance
 \cite{mewaldt,pdg,biermann,cronin,battistoni,Anchor} and references therein.

\section[Cosmic rays and...]{Cosmic rays and the natural background radiation}
\subsection{Introduction to cosmic rays}
Cosmic rays, first discovered by Victor Hess in 1912, are charged
particles accelerated at very high 
energies by astrophysical sources located anywhere beyond the
atmosphere of the Earth. 
$89$\% of cosmic rays consists of protons, followed by
$\alpha-$particles ($\sim 10$\%) and heavier nuclei ($\sim 1$\%) \footnote{
These data are taken from Ref.~\cite{mewaldt}. Of course, as it always
happens in the case of experimental data, there is some
uncertainty due to measurement errors. 
For this reason, different authors report slightly different
values as in the case of \cite{obrien}, which gives $95$\% for
protons, $3.5$\% for $\alpha$ particles and $1.5$\% for all the rest.
It should also be noted that the flux of particles below
the energy of $10$GeV strongly depends on the 11--year solar
cycle.}. 
All
elements of 
the periodic table, including  transuranic
elements, have been detected. 
More details on the
composition of 
cosmic rays can be found in Ref.~\cite{pdg} and references therein.

Within the flux of incoming particles one can
distinguish different components. The most relevant ones are 
 galactic CR (GCR), solar energetic
 particles (SCR) and the anomalous CR (ACR) \cite{nasaacr}. 
GCR 
originate from sources located in our galaxy outside the solar
system. 
It is believed that GCR
are a consequence of astrophysical events like stellar flares,
stellar coronal mass ejections, supernova explosions, particle acceleration
by pulsars
\cite{obrien}. 
Very often, the word cosmic rays refers only to GCR
\cite{mewaldt}. 
The smallest detectable energy of GCR is about 1~GeV. Below
this energy, the screening effect of the solar wind is too strong to
allow them to penetrate the heliosphere \cite{gapnote}. 
A detailed information on the interaction of GCR with the magnetic
fields of the Sun and of the Earth is contained in Ref.~\cite{obrien}
and references therein. Here 
we would just like  to mention the action on CR of the galactic
magnetic fields which are generated by the spinning of the Milky
Way. These fields are relatively weak, because the average magnetic
field in our galaxy is of the order of $10^{-10}$T. 
However, since CR consist of charged particles
traveling along huge distances, at the end these magnetic fields
are able to bend their trajectories in
a relevant way.
At this point one should  note that there are two
different components in the magnetic fields of our
galaxy, a regular one and a turbulent one, see
Ref.~\cite{RandKul}. The strengths and directions of the magnetic fields
belonging to the turbulent component are random
\cite{gapnote,RandKul}. Due to this 
randomness, the trajectories of CR are randomly bent. As a
consequence, the flux of CR  which arrives on the Earth is also random
or, more precisely, isotropic.
For this reason, it is
not easy to ascertain where GCR are coming from. Besides being
isotropic, the particle flux is
constant in time too, so that GCR form an almost uniform background of
ionizing radiation on the surface of the Earth. 
CR of energies up to  $10^{21}$eV have been observed.
These are considerable energies
for a microscopic particle. For example, the upper energy limit
of $10^{21}$eV corresponds in SI units approximately to 160 Joules. This is
comparable to the kinetic energy of
a ball of $0.8$kg thrown at the speed of
$50$km/hour. 
The origin of such ultra high energy CR (UHECR)
is so far unknown, but there are strong
hints that they could be produced outside of our galaxy. 
 Candidate sources of UHECR could be
relativistic 
plasma jets from 
supermassive black holes \cite{lofar}, explosions of galactic nuclei
\cite{obrien}, but other 
possibilities have been proposed, like magnetic monopoles,
see for example \cite{exotic}. 
The main evidence which suggests that UHECR are 
of extragalactic origin is provided by the fact that
the magnetic fields which are present in the Milky Way are
not able to 
trap CR of that energy.
Indeed, already protons of energy higher
than 10$^{15}$eV
are able to escape the galactic confinement\footnote{
In the case of heavier nuclei, the threshold energy for escaping the
galactic confinement is higher than that of protons. It is for this
reason that one observes a greater proportion of heavier nuclei with
respect to protons in CR with energy above 10$^{15}$eV.}.
As a consequence, if protons of energies of 10$^{19}$eV or higher
would be produced by sources located in our galaxy, they would escape
it in all possible directions following trajectories which are
almost straight lines. For this reason, the ultra high energy protons which
reach the Earth should arrive along directions which are approximately
parallel to the galactic plane. However, this conclusion is not
confirmed by observations, 
see Refs.~\cite{cseven,unscearb} and references
therein. Observations  show in fact
that the spatial
distribution of ultra high energy protons is isotropic, so that their
directions are not aligned with the galactic plane. This is of course
a strong hint that UHECR are of extragalactic origin.

On the other side, there is at least one argument which seems to point
out that the sources of UHECR are not very far from our galaxy.
In fact, it has been noted that
protons of energies above 5$\cdot$10$^{19}$eV 
would lose their energy by
interacting with the photons of the microwave (big bang)
background. 
This effect was predicted
in 1966, one year
 after the detection of the
  microwave background 
radiation, by Kenneth Greisen, Vadem Kuzmin and Georgi Zatsepin.
The energy threshold of 5$\cdot$10$^{19}$eV is called the GZK-limit,
from the names of its discoverers.
Protons with energies above that threshold will be slowed down
during their travel to the Earth by
the mechanism of energy-loss pointed out by Greisen, Kuzmin and Zatsepin
until their energy falls below the GZK-limit. This mechanism is
so efficient that, in practice,
protons with energies higher than 5$\cdot$10$^{19}$eV should not
be observed
on the Earth if their source is located at distances which are greater
than 50Mpc~\footnote{Mpc stands for megaparsec. 50Mpc are approximately 
150 millions of light years.
This is about 1500 times the diameter of a galaxy and it is
not a big distance in comparison with the cosmic scale of
distances.}. Since ultra high energy protons have 
instead been detected, this implies that they originate from
sources
which are within the range of 50Mpc. 
Yet the known cosmic objects which could be
able to accelerate
protons to such high energies are at least at a distance of about
100 Mpc or more.
So cosmic ray protons above this energy should not arrive on the
Earth
or we should explain why their formidable sources, which should be
relatively near to us, remain invisible.
These contradictions are known under the name of GZK
paradox.
On these points see Ref.~\cite{noelsta}.

Let's now discuss the remaining components of CR.
Energetic solar events, like for instance
solar flares, are able to 
accelerate particles up to some GeV very efficiently within the time
of 10
seconds. The SCR are mainly  
protons, heavier nuclei and electrons.
 Finally, it is worth  mentioning also the case of
ACR.
This kind of CR is mainly characterized by ions of elements which are
difficult to ionize, 
including He, N, O, Ne and Ar. Moreover, ACR have a relatively low
energy,
up to a few hundreds of MeV \cite{klecker}. It is thus improbable that
CR of such a low energy could originate  from the violent phenomena
which produce GCR. Besides, we recall that CR of energy below 1GeV
coming from outside the heliosphere cannot penetrate very deeply
inside the solar system, because they are deflected by the solar
magnetic fields.  There are indeed evidences that  ACR may be
neutrally charged dust
particles present in the interstellar gas near the border of the solar
  system \cite{mewaldt,nasaacr,jokipii}. When these particles
enter in contact 
with the far edges of the heliosphere, they are ionized by UV solar
photons or by interactions with the particles within the solar wind. 
Once these dust particles have been ionized, they are accelerated by
the shock waves formed when the solar wind encounters the interstellar
plasma. 
Particles escaping the shocks may diffuse toward the inner
heliosphere and arrive on the Earth as ACR. Actually, there are many
open questions on the mechanism of ACR acceleration. Hopefully, these
questions will be answered in 2007. By that date, in fact, the
Voyager~1 should reach the region in which the shocks
occur, which is thought to be somewhere
between 75 and 100 AU from the Sun
\footnote{At the time in which this article was
  finished, a distance up to about 90 AU has been explored by Voyager
  1. Some of the results of the observations can be found in
  Refs.~\cite{voy1,voy2,voy3}.}.  
This will be the first time that an example of
CR acceleration will be observed directly.

\subsection{Interaction of CR with the Earth's
  atmosphere}\label{subsec:intatm} 
When CR arrive near the Earth, they hit the nuclei of the atoms of the
 atmosphere, in particular nitrogen and oxygen, producing in this way secondary
particles.  The first interaction of the CR primary particle
takes place in the top 10\% of the atmosphere \cite{clay}. 
The most relevant reactions\footnote{One
should remember that the collisions of CR with the atmosphere gives
raise also to less relevant reactions, which
  produce  particles like  kaons, $\eta$
particles and even resonances.},  remembering that  
 90\% or more of CR consists of protons, are:
\begin{eqnarray}
pp\longrightarrow pn\pi^+\qquad &\mbox{or}&\qquad pp\longrightarrow
pp\pi^0\label{reaserone}\\
pn\longrightarrow pp\pi^-\qquad& \mbox{or}\qquad pn\longrightarrow
pn\pi^0\qquad \mbox{or}& \qquad pn\longrightarrow nn\pi^+\label{reasertwo}\\
\end{eqnarray}
In the above reactions all the secondary particles are hadrons, namely
protons $p$, neutrons $n$ and pions in all their charged states
$\pi^\pm,\pi^0$. 
Pions may in turn decay according to the following processes:
\begin{eqnarray}
\pi^+\longrightarrow \mu^+\nu_\mu&\mbox{and}&
\pi^-\longrightarrow\mu^-\bar \nu_\mu\label{decaserone}\\
\pi^0&\longrightarrow&\gamma\gamma\label{decasertwo}
\end{eqnarray}
where the $\mu^\pm$'s are muons,
the $\gamma$'s are photons and $\nu_\mu,\bar \nu_\mu $ are
respectively muonic
neutrinos and their anti-particles. 
The mean life of pions 
is 26ns for $\pi^\pm$ and 10$^{-16}$ seconds in the case of
$\pi^0$. For this 
reason, charged pions may 
still collide with air atoms before decaying, but it is very
unlikely that this happens in the case of
neutral pions, which have a very short average life.
Other secondary particles, like protons, neutrons  and photons
 interact
very frequently with the atoms of the atmosphere 
giving raise in this way to a cascade of less and less energetic secondary
particles. At the end, these particles are  stopped 
by the atmosphere or, if the energy of the primary particle
was sufficiently high, they can reach the ground.

\begin{figure}[bpht]
\centering
\includegraphics[width=\textwidth]{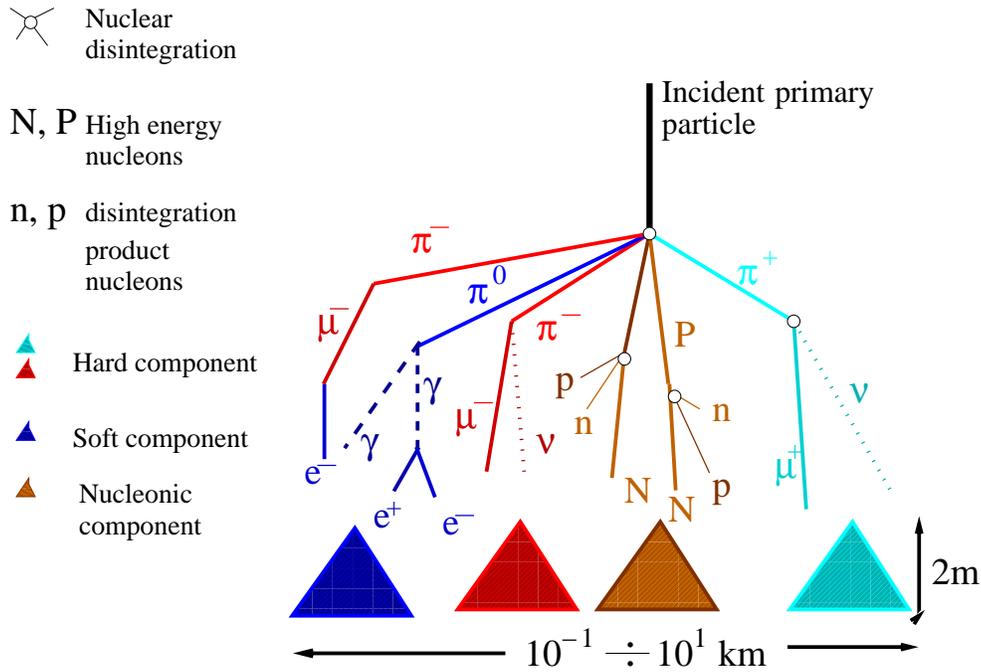}
\caption{This figure illustrates schematically how air showers
  generate from cosmic rays. The high energetic  primary particle,
  usually a proton, 
  whose trajectory has been denoted in black,
  starts to interact with the molecules of the upper atmosphere. In this
  way, secondary particles are
  produced, which give raise to other particles (tertiary, quaternary
  etc.) via other interactions with the atmosphere or via decay
  processes. The total flux of particles can be divided in a {\it
  electromagnetic
  component} (photons, electrons and anti-electrons or positrons), see the
  blue trajectories in the figure, in a {\it muon component}, see the
  cyan and red trajectories, and finally in a {\it nucleonic component}
  (mainly protons, neutrons, rarely pions)
  denoted in brown. 
  At the sea level, the air shower has the form of a pancake, whose
  height is around two meters, while its radius usually around a few
  hundreds of meters, but may reach some tents of kilometers in the
  case of very energetic cosmic rays. 
  The nucleonic component is usually confined in a
  narrow cone centered along the direction of the incoming primary
  particle. 
}\label{shower}
\end{figure}

The main mechanism of energy loss \footnote{Here and
  in the following energy means the 
  kinetic energy of the particles.} of high energetic hadrons
consists in the disintegration of the molecules of the atmosphere
\cite{milagro}, 
see Fig~\ref{shower}.  This leads to the creation of new
particles through nuclear interactions 
  like those shown in Eqs.~\ceq{reaserone} and \ceq{reasertwo}.
At lower energies,
   dissipative processes become predominant, in which 
the molecules of the atmosphere get either ionized or excited.
The most relevant process of
  this kind in the case of heavy charged particles is the ionization
  of the molecules of the atmosphere. Lighter charged particles like
  electrons and positrons lose their energies not only by ionization,
  but also by 
  bremsstrahlung. This consists in the
  radiative loss of energy of charged particles moving inside matter,
  when they are deflected by the electrostatic forces of the positive
  charged nuclei of the surrounding molecules. Photons and neutrons,
  the remaining  relevant particles in the cascade,
  are examples of indirectly ionizing radiation. Their interaction
  mechanisms are more complicated than those of charged
  particles and will not be described here. The interested reader may
  find a more detailed account on the way in which radiation of
  different kinds interacts with matter in Ref.~\cite{doereport}.

The total number of secondary particles $N_{sec}$ within the
cascade grows rapidly, mainly sustained by the processes of
bremsstrahlung and pair production due to electrons, positrons and
photons. Hadrons like  protons and, to a less extent, neutrons,
are easily stopped 
by the atmosphere, so that they
increase relevantly the number of particles by
disintegrating the 
molecules of air only during the first stages of the formation of the
cascade.
The
decays of pions given in Eqs.~\ceq{decaserone} and
\ceq{decasertwo} produce  muons and
photons of considerable energies.
The muons are very penetrating particles and do not interact very much
with the air. They lose a small fraction of their
energy before reaching the ground by ionizing the molecules of the
atmosphere. 
The photons give raise instead to electron-positron pairs
$e^+e^-$. In turn, electrons and positrons create other electrons by
ionization or other photons due to bremsstrahlung.
In this way, while the cascade propagates inside the atmosphere,
the number of its electrons, positrons and photons grows almost
exponentially. 
The maximum number of  particles inside the cascade is
attained when the average energy per electron reaches the
threshold $E_T\sim$80MeV. When the energy of electrons in air falls
below that threshold,
ionization starts
to prevail over bremsstrahlung as the main mechanism of energy loss of
electrons in air
and the process for increasing the number of particles 
described above ceases to be effective.

If the  
energy of the primary
particle is below $10^{14}$eV, essentially only the penetrating muons and 
neutrinos are able to arrive to the sea level, while the
other particles in the cascade are absorbed at higher altitudes.
Actually, neutrinos interact so rarely with matter that they could
pass through a light year of water without undergoing any
interaction. Thus, if one is concerned with the dose of ionizing
radiation delivered by CR to the population, the contribution of
neutrinos  can 
simply be neglected. Muons are more dangerous for the health.
They have a short
mean life at rest (2.2ns), but since they travel at very high speeds,
they manage to reach the surface of the Earth due to the relativistic
dilatation of 
time. Part of these muons can still decay, giving raise to 
electrons $e^-$ or
positrons $e^+$, mainly according to the process:
$\mu^\pm=e^\pm+\nu_e+\nu_\mu$.

When the energy of the primary particles is instead above $10^{14}$eV, the
cascade of secondary particles arrives to the ground before being
stopped by the atmosphere. In that case, the cascade is called with
the name of air
shower, see Fig.\ref{shower}.
To be precise, the effects of an air shower started by a primary
particle of 10$^{14}$eV are relevant up to altitudes comparable to
that of Mount Everest \cite{allan,lofara}.
Only air showers generated by primaries of energy
of about $10^{15}$eV or higher are
able to reach also the typical altitudes of inhabited areas and arrive
to the
sea level. The frequency of these air showers is relatively high,
because the  total flux of primary particles with energy
$E\ge 10^{15}$eV is of about 100 particles per m$^2$ per year.
Giant air showers produced by primaries of energies
beyond $10^{20}$eV are much more rare: their total  flux
is of 1 particle per km$^2$ per century \cite{mewaldt}.
More data concerning fluxes of incoming primary particles and an
explanation of how these data are measured using ground detectors, can
be found in Refs.~\cite{Anchor,eheas}.

In the air shower we distinguish a nucleonic component, a muon
component and an electromagnetic component. The nucleonic
component is generated by
high energetic protons and neutrons, which  disintegrate the
atoms of the air giving raise to other
protons and neutrons. 
The fluxes of electrons, positrons and photons started  by the
decay of the $\pi^0$'s,  together with the electrons
and positrons 
coming from the decay of muons or from the ionization due to the hadrons, form
the electromagnetic component.
In the early times of CR research, electrons and positrons
have been called the soft component, while
 muons coming from the decay of the charged pions have been
called the hard component. These names originate from the fact
that muons are very penetrating and thus they may be regarded as ``hard''
particles. For instance, at the energy of 1 GeV, the 
range\footnote{The range is defined as
the average depth of penetration
of a charged particle into a material before it loses all its kinetic
energy and stops. The concept of range has a meaning only in the case
of charged particles whose energy is kinetic energy which is
lost continuously along their paths due to ionization and
bremsstrahlung processes
\cite{doereport}.} of muons
is  2.45$\cdot$10$^{5}$g~cm$^{-2}$. This implies
that in water, which has a density of 1~g~cm$^{-3}$ at 4~C, muons
run along an average distance of 2.45 km before being stopped
completely. In standard rock, which has a density of 2.65~g~cm$^{-3}$
\cite{pdg}, this average distance reduces to about 900 meters.

At the sea level, the air shower has approximately the form of a
pancake with an height of 1-2  meters.  
Its extension in the other two directions, defined
as the distance in which 90\% of the total energy of the shower is
contained, is given by the so-called Moli\`ere radius. 
For example, in the case of an air shower started by a primary
particle of an 
energy of $10^{19}$eV (10 
EeV) 
the Moli\`ere radius  has a length of about 70 meters.
The real extension of the
shower is  much bigger and some of the muons may be detected
up to a distance of a few kilometers from the core \cite{eheas}.
Usually the nucleonic component, which is composed by heavier
particles than those of the muon and electromagnetic components, is
less deflected 
 from the
direction of  the incident primary particle
by the
interactions
with the atmosphere  and it is thus concentrated in a narrow
cone inside the air shower. The center of the cone is roughly aligned
with the direction of the original primary particle. 
The number of secondary particles which 
arrives on the ground with an air shower 
is huge. Considering particles whose energies
are greater that 200keV, an air shower generated by a 10EeV primary
particle contains up to 10$^{10}$ particles, mostly photons,
electrons 
and positrons. Electrons outnumber positrons in the ratio 6 to 1.
The maximum number of
particles, i. e. the so-called point of shower maximum or simply shower
maximum, is attained at an altitude of 2--3 km above the sea
level. 
 Many other data and diagrams describing the
propagation of air 
showers in the atmosphere may be found in Ref.~\cite{TalkPierog}.

\begin{figure}[bpht]
\centering
\includegraphics[width=10cm]{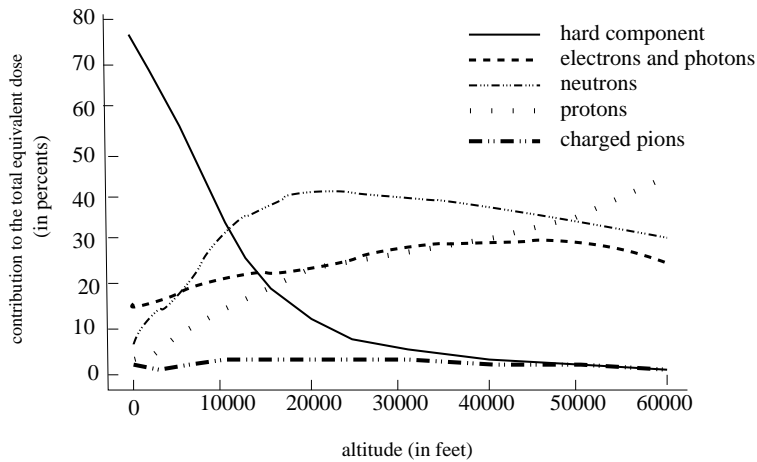}
\caption{Percent contribution of the various CR components to the
  total equivalent dose at different altitudes. This figure is based
  on data of Ref.~\cite{obrien}.}\label{percent}
\end{figure}

When air showers approach the ground, about 85\% of their energy is
concentrated in the electromagnetic component. The contribution of the
muon and nucleonic components is thus much less relevant. 
 The situation changes completely if we consider all CR, and not only
those which have sufficient energy to give raise to an air shower.
As we see in Fig.~\ref{percent}, muons are in fact responsible
for about 85\% of the total equivalent dose delivered by CR to the population
at the sea level. 
As a consequence, globally it is the muon component the most significant
from the energetic point of view and not the
electromagnetic component. 
The reason of this fact is that primary particles with energy $E\ge
10^{15}$, namely those which can 
produce air showers, form just a minimal fraction of the total
amount of CR arriving on the Earth.
For example, the 
total flux of particles with energy $E\ge 10^{12}$eV (1 TeV), is of
1~particle~per~m$^2$~per~second, i. e. a factor of 3$\cdot10^5$
higher than the  total flux of CR with energy $E\ge
10^{15}$ reported above.
In other words, there is an overwhelming number of CR with energy lower than
$10^{15}$eV which are not able to start an air shower, but may
still generate energetic muons. Being very penetrating
particles, these muons are not stopped easily by the
atmosphere and reach the sea level altitude, where they
represent the biggest source of ionizing radiation of cosmic origin.
Other particles which deliver relevant doses of ionizing
radiation to the population on the surface of the
Earth
 are photons, electrons and
neutrons.  The percent contributions to the total equivalent dose of the
various  components of CR as a function of the altitude is given in
Fig.~\ref{percent}. In that figure, which has been published in 1996,
 the curve
concerning neutrons should be taken with some care, because
 the data on neutron fluxes in
the atmosphere were still sparse at that time \cite{unscearb}. Other
data about energies and fluxes 
of particles due 
to CR will be given in the next Section.
We note that protons and neutrons prevail at higher altitudes, but
they are rapidly absorbed by the atmosphere and
muons become dominant at lower altitudes.

\subsection{Intensities and fluxes of CR}
CR are the source of an avalanche of secondary particles which hit
constantly the surface of the Earth. To determine the energy and
number of these particles, their directions of arrival and their
distribution in time, quantities like the integral vertical intensity
or the differential directional intensity are measured. The meaning of
these quantities is explained in details in a separate
Appendix at the end of this article. In this section, we present some
experimental data which are 
useful in order to characterize the contribution of the muon,
electromagnetic and nucleonic components to the background of ionizing
radiations on the ground due to CR.

\begin{figure}[bpht]
\centering
\includegraphics[width=10cm]{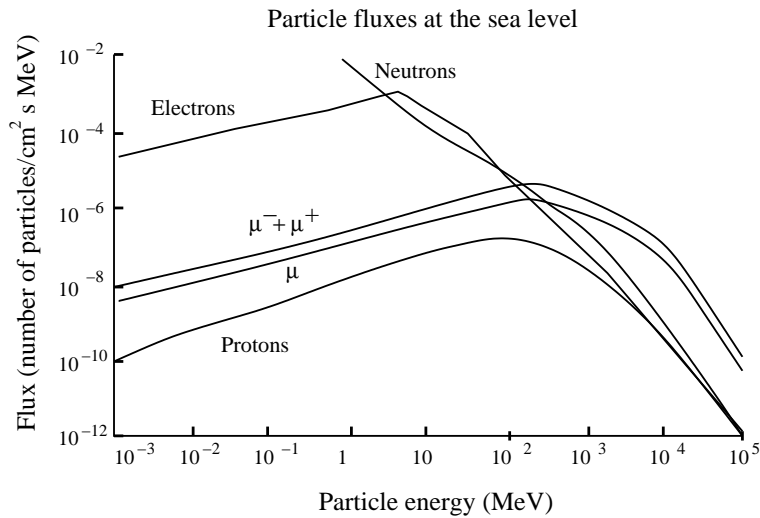}
\caption{Differential integrated flux (see Eq.~\ceq{fluangthe} of the
  Appendix)
of the different components of
  CR related 
  radiation at the sea level. This figure has been created on the
  basis of an analogous figure appeared in Ref.~\cite{schumacher}. The
  original data are
taken from
Ref.~\cite{ziegler}.}\label{ziegler}
\end{figure}

The  integral vertical intensity (IVI in the Appendix,
see Eq.~\ceq{ividef}) of the muon component
with energy above 1 GeV at sea level is approximately
$I_{ivi}^{hard}(\theta=0)\sim$0.70$\cdot 10^{-2}$
cm$^{-2}$s$^{-1}$sr$^{-1}$\cite{pdg}. 
The integral directional intensity of muons in the other
directions, which are at an angle $\theta$ with respect to the vertical
direction, has the following behavior:
$I_{idi}^{hard}(\theta)\propto I_{ivi}^{hard}(\theta=0)\cos^2\theta $.
More complete phenomenological formulas for the angular distribution of CR
intensities may be found in Refs.~\cite{pdg,allkofer,dar}.
As Fig.~\ref{ziegler} shows,  muons arriving to the ground are very
energetic. The most frequent muon energy is 500MeV, while the
average muon energy is 4GeV. 
There is almost no protection from this
source of radiation, since, as we have seen before,
 high energetic muons are able to
penetrate thick layers of concrete and rocks.

\begin{figure}[bpht]
\centering
\includegraphics[width=10cm]{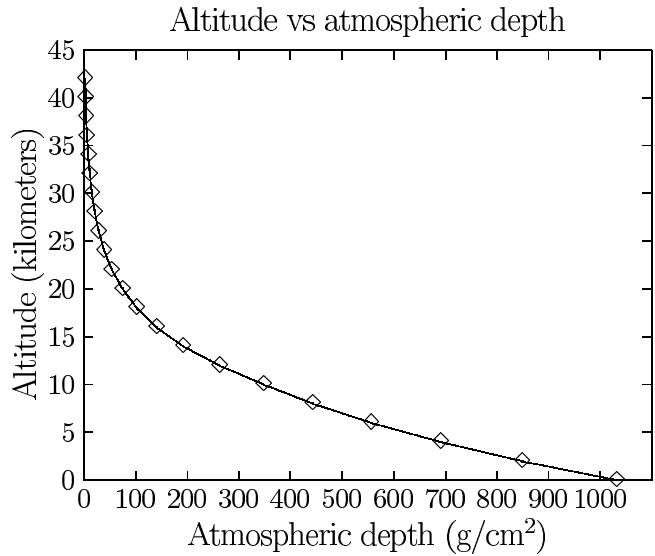}
\caption{Conversion diagram from atmospheric depth to altitude. The
data corresponding to the dots are taken from
Ref.~\cite{usat}.}\label{altdep} 
\end{figure}

The integral vertical intensity for electrons plus positrons with
energies greater than 10, 100 and 1000 MeV is very approximately given
by\cite{pdg}:
\begin{eqnarray}
I_{ivi}^{el+pos}(\theta=0, E_{min}>10MeV)&\sim& 0.30\cdot 10^{-2}
\mbox{cm}^{-2}\mbox{s}^{-1}\mbox{sr}^{-1}\\
I_{ivi}^{el+pos}(\theta=0, E_{min}>100MeV)&\sim& 0.06\cdot 10^{-2}
\mbox{cm}^{-2}\mbox{s}^{-1}\mbox{sr}^{-1}\\
I_{ivi}^{el+pos}(\theta=0, E_{min}>1000MeV)&\sim& 0.02\cdot 10^{-3}
\mbox{cm}^{-2}\mbox{s}^{-1}\mbox{sr}^{-1}
\end{eqnarray} 
Moreover, the total flux of electrons plus positrons amounts
approximately to 30{\tt \char`\%} of the total particle flux 
which is reaching the ground
due to cosmic
rays. Always according to \cite{pdg}, 
the ratios of photons to electrons + positrons is approximately 1.3
above 1 GeV and 1.7 below the critical energy $E_T\sim 80$MeV mentioned above.
The differential flux on the ground of the various components of 
radiation related to CR can be seen in Fig.~\ref{ziegler}.
One may observe that
different particles 
become predominant at different energies. 
In the lowest portion of the energy range,
 electrons are
predominant. At energies of around 1~MeV electrons are taken over by
fast  neutrons which arise mainly due to the de-excitation of
atmospheric nuclei following compound-nucleus reactions
\cite{desalin} \footnote{Roughly speaking, in a compound nucleus
  reaction a neutron or a proton, but also an $\alpha-$particle
  interacting with the nucleus of an 
  atom creates a nucleus of higher atomic number which is metastable
  and decays after a short period of time.
For example, a possible compound-nucleus reaction is:
$p+{}^{63}Cu\longrightarrow{}^{64}Zn^*$. The de-excitation of the
metastable nucleus  produces others neutrons and protons, e. g.:
${}^{64}Zn^*\longrightarrow {}^{63}Zn+n$ or
${}^{64}Zn^*\longrightarrow {}^{62}Cu+n+p $. Compound-nucleus
reactions become possible only if the energies of the incoming
nucleons or alpha particles are such that the de Broglie wave lengths
of these particles
are comparable with the size of the hit nucleus.}.
Electrons become once again predominant in the energy range going
from a few MeV up to some tens of MeV. Starting from
energies approximately above 200MeV, the number of muons becomes
overwhelmingly high in comparison to that of the other particles.
In considering the above data, one should of course take into account
the fact that, at low energies, let's say
below 10MeV,
particle fluxes are strongly
dependent on many factors, including weather conditions, so that there are
big uncertainties in their measurement up to an order of magnitude
\cite{ziegler}. 
Moreover, all the  data presented so far in this Section
refer to the sea level
altitude. With 
increasing altitudes, 
the contribution to the particle flux given by protons and neutrons, the
nucleonic component of CR, becomes more and more relevant (see
Fig.~\ref{percent}) and is
predominant above atmospheric depths   of approximately
500~g~cm$^{-2}$~\footnote{The atmospheric depth is a quantity which is
  often used to measure the altitude. For convenience, we give in
  Fig.~{\ref{altdep}} a diagram which is useful to make the conversion from
  atmospheric depth in g/cm$^2$ to altitude in kilometers.
}. 

To conclude this Section, we provide also some data regarding the particle
flux outside 
the heliosphere.  The total flux of CR in the
galaxy is large, about
100000 particles m$^{-2}$s$^{-1}$ \cite{zieglerb}.
Much lower is for example the integral directional
flux of CR primaries with $E>2\cdot
10^{15}$eV is about $\Phi_{idf}=75000$ particles
km$^{-2}$sr$^{-1}$day$^{-1}$. 
This datum has been derived using a formula for the integral
directional flux given in Ref.~\cite{lofara}, which is based on the
results of measurements and simulations of incoming cosmic ray fluxes
reported in Ref.~\cite{fowler}.
The dependence of this flux
on the direction of the
incoming particles is minimal since, as mentioned before, the
distribution of these particles is isotropic due to the presence of
random  magnetic fields in the Milky Way. 
 Finally, the integral directional flux of particles with energies
above $10^{20}$eV is
$\Phi_{idf}=1 \mbox{ particle}
\mbox{ km}^{-2}\mbox{ sr}^{-1}\mbox{ century}^{-1}$ 
\cite{kitpmini}. With such a small flux, the investigation of
CR with energy beyond the GZK limit requires detectors which cover a
very large area. Perhaps with the next generation of detectors it
will be possible to solve the puzzles connected with UHECR.

\subsection{Dose of ionizing radiation from CR in present times}
In present times\footnote{The data concerning the present levels of
  radiation coming from CR are taken from the United Nations Report of
  the year 2000
  \cite{unscearb}. } the effective dose rate (EDR) delivered by CR to the
human population
varies
from a minimum of 300$\mu$Sv~year$^{-1}$ to a maximum of
2000$\mu$Sv~year$^{-1}$.
This wide range depends on many factors, the most important one being
the altitude. At sea level, the value which usually is given for the EDR is
270~$\mu$Sv~year$^{-1}$, or equivalently 31~nSv~h$^{-1}$.
 One should
keep in mind however that this estimation is the result of
population-weighing average. As a matter of fact, even if the
altitude is fixed at the sea level,
the EDR still changes with the latitude within a range of
variation of approximately 10\%. 
Strictly speaking, 
270~$\mu$Sv~year$^{-1}$ 
is the dose rate
received by the population living at a latitude
which is near the 30$^\circ$ parallel. 
It turns out in fact from the distribution of the
population on the Earth that this is the
average latitude at which people are living. Besides, the above value
of EDR takes into account only the contributions of  muons and of the
electromagnetic component. 
The nucleonic component, which at the sea level is essentially
consisting of neutrons, gives 
to the
average EDR at the sea level an additional contribution of 
48~$\mu$Sv~year$^{-1}$
or, equivalently, 
5.5~nSv~h$^{-1}$. The figures concerning neutrons
 should be taken with some care, since
up to the time in which Ref.~\cite{unscearb} was released,
 the available data on neutron fluxes were sparse. If one considers
 also the different altitudes in which the population lives, the 
population-weighted EDR is of 380~$\mu$Sv~year$^{-1}$, corresponding
  to an average habitant of the planet Earth living approximately near the
  30$^\circ$ parallel and at an altitude of 900 meters above the sea
  level. 
\section{Conclusions}
In this review a short but thorough account has been provided on what
it is known about cosmic rays, starting from their origin in space and
arriving to the doses of ionizing radiation
delivered by them to the human population.
Only the sources of CR have not been discussed here, because this
argument is outside the aims of this article.
Much attention has been dedicated to fluxes and intensities 
of CR and of the particles arriving on the ground as an effect of the
cascades  initiated by CR in the atmosphere.
These quantities are  of interest for scientists working in
different subjects. Apart from
 research in high energy physics and astronomy,
the fluxes and intensities of particles of cosmic origin are also
studied  for radioprotection purposes \cite{obrien,unscearb} and 
for their 
capability of causing potentially harmful failures in computers and
electronic storage devices \cite{zieglerb}.
Fluxes of CR  are also carefully measured due to their
relevance to space explorations, see for example \cite{ssareport}.
Finally, in  Appendix A
 the various kinds of fluxes and intensities of CR and related
particles which one encounters in research articles about CR  have
been defined and their meaning has been illustrated.
Concrete expressions for these quantities have been given in
terms of
mass densities, velocity distributions  and energy densities. Both
relativistic and non-relativistic cases have been treated. Up to
now, a systematic classification and explanation of these quantities
such as that provided in this work was 
missing in the scientific literature on CR. The necessity of filling
this gap justifies the length of this Appendix.
\begin{appendix}
\section[Appendix A: Definitions...]{Appendix A: Definitions of
  intensity, flux and related quantities}

We have seen that, in order to
characterize the intensity and flux
of  charged particles which arrive on the
Earth due to CR
there exist an entire zoo of observables.
Their names and meanings may sound puzzling for somebody who
is not acquainted with them.
Moreover, the
same observable is sometimes called with different names by different
authors, or, on the contrary, the same 
name is used to describe two slightly different observables in different
contexts. Besides, it is not easy to find an explanation of these
observables in the scientific literature. Books on radiative
transport contain often useful information, see for example
\cite{ryblig}, however these books describe the intensity and flux of
radiation emitted by an energy source. Here we are instead dealing
with intensity and flux of particles arriving at a detector.
For these reasons and also to 
make this article self-contained,
in the following it will be made an effort to explain the meaning of the
various quantities which are relevant in the
physics of CR. 

\subsection{Differential directional intensity}\label{subsec:aaaa}
 The
  differential directional intensity (DDI) $I_{ddi}$ is defined
  \footnote{Here we 
  follow Ref.~\cite{rossi}, in which a very clear and precise definition
   of the related concept of directional intensity is presented}
in such a way that the quantity 
\begin{equation}
dN_i= I_{ddi}dSd\Omega dE_i
  dt\label{defddi} 
\end{equation}
 represents the
  number of particles of a given kind incident upon the infinitesimal
  element of 
  area $dS$ during the time $dt$ within the element of solid angle
  $d\Omega$ perpendicular to $dS$ and within the energy interval $[E_i,
  E_i+dE_i]$. Here the index $i=1,2,3,\ldots$ labels the different kinds
  of particles (electrons, protons, muons etc.)

\begin{figure}[bpht]
\centering
\includegraphics[width=10cm]{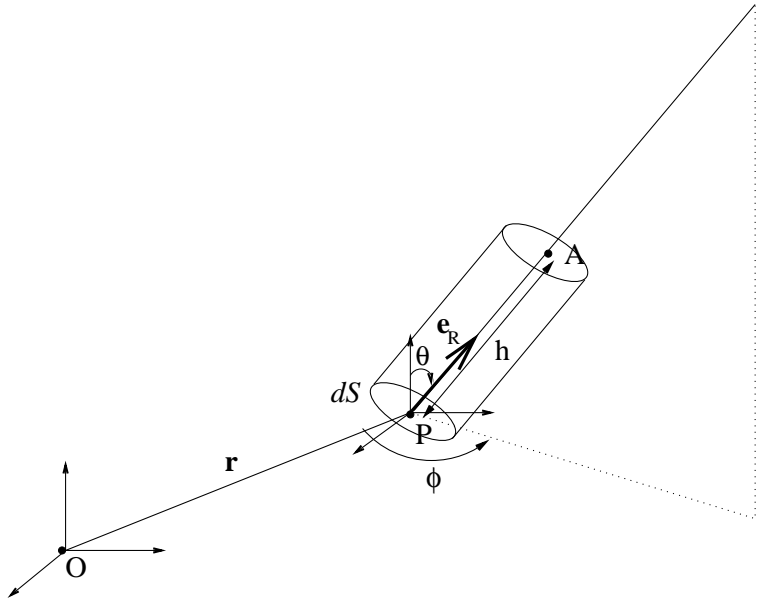}
\caption{This figure shows the geometrical setup for the definition of
  the differential directional intensity. The normal vector  $\mathbf
  n$ to the 
  infinitesimal surface $dS$ at the point $P$ coincides with the
  vector $\mathbf e_R$ which gives the direction of the incoming particles.
The particles that will be traversing the surface $dS$ within the interval
  of time $dt$ are those contained in the volume $h\cdot dS$, where
  $h=|\mathbf v_i|dt$.
}\label{dirintensity}
\end{figure}

To compute explicitly the $I_{ddi}$ in terms of physical
 parameters like particle velocity and mass or energy density, let us
 consider a point 
 $P$ in the space, whose position with respect to a cartesian system of
 coordinates $Oxyz$ is given by the radius vector $\mathbf
 r=(x,y,z)$. 
In the following, it will be convenient to define a second
 reference system with origin in $P$ and 
 spherical coordinates $\varpi,\theta, \phi$, with $0\le\theta\le \pi$
 and $0\le\phi \le 2\pi$, see Fig.~\ref{dirintensity}. 
We introduce also the infinitesimal vector element of surface $d\mathbf
 S=dS\mathbf n$. The area and the
 orientation of $d\mathbf S$ are given 
 by $dS$ and by the unit vector $\mathbf n$ which is normal to $dS$
 respectively. The element of surface $dS$ is centered around the point $P$.
Now we wish to count the number of particles of a certain
 type, e. g. electrons, which hit per unit of time
 the surface $d S$ and whose velocities $\mathbf v_i$ are
 oriented according 
 to  a certain direction, given for instance by the
 unit vector 
$\mathbf e_R(\theta,\phi)$.
In mathematical terms this last condition is expressed as follows:
$\mathbf v_i=
|\mathbf v_i| \mathbf e_R(\theta,\phi)
$.
We will see that, in the case of the flux, the direction of
$\mathbf e_R(\theta,\phi)$ may be arbitrary. However, in the case of the
 intensity, the element of
 surface
$d\mathbf S$
should be by definition perpendicular to 
the vector  $\mathbf e_R(\theta,\phi)$. In other
words:
\begin{equation}
\mathbf n=
\mathbf e_R(\theta,\phi)\label{norsurfcond}
\end{equation}
If the particles are non-relativistic, one may express the DDI
in terms of the velocity and  mass
density of particles:
\begin{equation}
\mathbf v_i=|\mathbf v_i(E_i)
|\mathbf e_R(\theta,\phi)
\label{fdrfe}
\end{equation}
\begin{equation}
\rho_i=\rho_i(\mathbf r,\theta,\phi,E_i,t)\label{massdensity}
\end{equation}
 The norm of $\mathbf v_i$ is  a function of the energy of the
particle given by the
 well known relation:
\begin{equation}
 E_i=\frac{m_i}2|\mathbf v_i|^2 \label{kinene}
\end{equation} 
Let us note that the distribution of
density of mass $\rho_i$, which can also be a sum of Dirac delta
functions, depends on the angles $\theta,\phi$, on the
position $\mathbf r$ of the point $P$ in the space, on the energy
$E_i$ and on the time. 
There is however no
dependence on the radial coordinate $\varpi$.
We will see below why it is not necessary to add the radial coordinate
in the list of the arguments of $\rho_i$.
Looking at Fig.~\ref{dirintensity}, it is clear that the number of
particles incident upon the surface $dS$ from the specified
direction is 
given by:
\begin{equation}
dN_{i,\mathbf e_R}(\mathbf r,\theta,\phi,E_i,t)=
\frac{\rho_i}{m_i}|\mathbf v_i| d\mathbf S\cdot
\mathbf e_R(\theta,\phi)
dt\label{eeerer}
\end{equation}
In order to derive Eq.~\ceq{eeerer} it has been used the fact that the
total mass $dM$ of the particles which traverse the surface $dS$ in the
 interval of time $dt$ is  $dM=\rho_i|\mathbf v_i|
 d\mathbf S\cdot \mathbf e_R dt$. The  number of such particles
is  obtained
 after dividing the total 
 mass $dM$ 
by the mass $m_i$ of a single particle of type $i$.
The dependence of
$dN_{i,\mathbf e_R}(\mathbf r,\theta,\phi,E_i,t)$
 on the kinetic energy $E_i$ becomes explicit after eliminating the
 norm of the velocity $|\mathbf v_i|$ from the
 right hand side of Eq.~(\ref{eeerer}) using Eq.~(\ref{kinene}).
 Finally, the scalar product $d\mathbf S\cdot
\mathbf e_R(\theta,\phi) $ gives the effective area which is hit by
the particles incoming from the direction $\mathbf
e_R(\theta,\phi)$. Since by definition of DDI 
the particle velocities are
always perpendicular to $d S$, i. e. parallel to $d\mathbf S$,
see Eq.~(\ref{norsurfcond}),
we have that
\begin{equation}
dN_{i,\mathbf e_R}(\mathbf r,\theta,\phi,E_i,t)=
\frac{\rho_i}{m_i}|\mathbf v_i| dS
dt=\frac{\rho_i}{m_i}\sqrt{\frac{2E_i}{m_i}}dSdt
\label{numparunitsolidangle}
\end{equation}
Now we are in a position to understand why it is not needed to take
 into account 
 the dependence on $\varpi$ of mass density.
The reason is that just a small portion of space near the
 point $P$ is considered, so that the radial coordinate
is varying within the
interval $[0,h]$, where
 $h=|\mathbf v_i|dt$. Clearly,
the variation of $\rho_i$
with 
 respect to  the radial coordinate is negligible within this
 infinitesimal interval.

\begin{figure}[bpht]
\centering
\includegraphics[width=8cm]{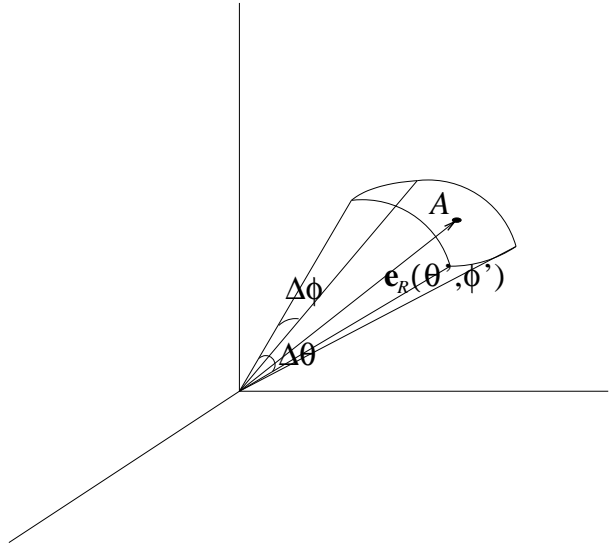}
\caption{Area $A$ spanned on a sphere of unit radius
 by the unit vectors $\mathbf
  e_R(\theta',\phi')$. The values of $\theta'$ and $\phi'$ are defined
 in Eqs.~\ceq{intone} and \ceq{inttwo}.
}\label{areaspaner}
\end{figure}

Let us remark that, in  real measurements, the number of
particles coming from  
a particular
 direction is usually very small, so that
 it is better to consider an entire set
 of directions, for instance those characterized by slightly different
 angles $\theta',\phi'$ included within the range
\begin{eqnarray}
 \theta\le\theta'\le\theta+\Delta\theta\label{intone}\\ 
 \phi\le\phi'\le\phi+\Delta\phi\label{inttwo}
\end{eqnarray}
where $\Delta\theta$ and $\Delta \phi$ denote finite
 quantities and not 
 infinitesimal ones.
Clearly, the unit vectors  $\mathbf e_R(\theta',\phi')$ associated to
 these directions span a surface of area:
\begin{equation}
A=\int_\theta^{\theta+\Delta\theta}d\theta'\sin(\theta')
\int_{\phi}^{\phi+\Delta\phi}d\phi'\label{areunisph}
\end{equation}
on a
 sphere of unit radius, see Fig.~\ref{areaspaner}.
Always for experimental reasons, it will also be convenient to
enlarge the set of possible particle energies to a finite interval:
\begin{equation}
E_i\le E'_i\le E_i+\Delta E_i\label{intene}
\end{equation}
Now we count the incoming particles in the neighborhood of the point
$P$
according to the following
procedure.
For each value of the energy $E_i'$ within the interval (\ref{intene})
 and
for each of the  directions
falling within the range  of angles of
Eqs.~(\ref{intone}--\ref{inttwo}), we count the number of particles
traversing a surface 
 of fixed area $dS$ and perpendicular to that direction in the sense
 of Eq.~(\ref{norsurfcond}). Successively, summing over all the numbers
 obtained in this way, we  get the final result
 $dN_{i,\Delta\theta,\Delta\phi,\Delta E_i}(\mathbf
r,\theta,\phi,E_i,t)dSdt$.
In terms of mathematical equations, this means that the quantity 
 $dN_{i,\Delta\theta,\Delta\phi,\Delta E_i}(\mathbf
r,\theta,\phi,E_i,t)$
is computed by integrating the
infinitesimal number of particles $dN_{i,\mathbf e_R}(\mathbf
 r,\theta,\phi,E_i,t)$ of Eq.~(\ref{eeerer})
as follows:
\begin{equation}
dN_{i,\Delta\theta,\Delta\phi,\Delta E_i}(\mathbf r,\theta,\phi,E_i,t)=
\int_{E_i}^{E_i+\Delta E_i}
\int_\theta^{\theta+\Delta\theta}
\int_{\phi}^{\phi+\Delta\phi}
\frac{\rho_i}{(m_i)^\frac
  32}\sqrt{2E_i} d\Omega'
dSdt\label{eeererint}
\end{equation}
To write the above equation we have used Eq.~(\ref{kinene}) in order
 to express the 
 speed $|\mathbf v_i|$ as a function of the kinetic energy $E_i$ and
 the fact that the infinitesimal element of solid angle $d\Omega'$ is
 given by:
\begin{equation}
d\Omega'=\sin(\theta')d\theta'd\phi'\label{solangdef}
\end{equation}
 The quantity $dN_{i,\Delta\theta,\Delta\phi,\Delta E_i}(\mathbf
 r,\theta,\phi,E_i,t)$ describes 
 the intensity of particles with energy in the interval (\ref{intene})
which arrive at the point $P$ from all the
 directions spanning the area $A$
of Eq.~(\ref{areunisph}) on a sphere of unit radius.

To compute the DDI, it is now sufficient to
take the limit in
 which $\Delta\theta$ and $\Delta\phi$ become
infinitesimally small. In that case:
\begin{equation}
dN_{i,\Delta\theta,\Delta\phi,\Delta E_i}(\mathbf 
r,\theta,\phi,E_i,t)
\sim dN_i(\mathbf r,\theta,\phi,E_i,t)
\end{equation}
with
\begin{equation}
dN_i(\mathbf r,\theta,\phi,E_i,t)=
\frac{\rho_i}{(m_i)^\frac
  32}\sqrt{2E_i} 
d\Omega dE_i dSdt\label{gee}
\end{equation}
It is easy to realize that the number of particles
$dN_i(\mathbf r,\theta,\phi,E_i,t)$ of Eq.~(\ref{gee}) coincides with
the number of  particles entering in the definition of DDI of
Eq.~(\ref{defddi}). 
Comparing these two equations, we find that
\begin{equation}
I_{ddi}(\mathbf r,\theta,\phi,E_i,t)=
\frac{\rho_i}{(m_i)^\frac
  32}\sqrt{2E_i} \label{ddiexplfinal}
\end{equation}
Eq.~(\ref{ddiexplfinal}) provides a nice relation between the DDI and
the mass density $\rho_i$.
From Eq.~(\ref{ddiexplfinal}) it turns out that the
 units in which the DDI  is measured are cm$^{-2}$s$^{-1}$sr$^{-1}$GeV$^{-1}$,
where sr is 
a shorthand for steradiant, the unit of solid angles. In SI units one
should replace centimeters by meters and GeV's by Joules.

As a final remark, let us note that the
DDI carries a lot of information about the intensity of particles
incoming from different directions. We will also see that the DDI is
related to the energy density of particles.
On the other side, 
the
DDI 
is an observable which is mainly  related to the
 point in which it is measured. In particular, it is
 not possible to integrate the quantity $I_{ddi}$ in $dS$
in order to extend its meaning to an arbitrary
 finite surface $S$. 
The reason is that, in the definition of the DDI the vector element of
surface $d\mathbf S$ is constrained to satisfy
Eq.~(\ref{norsurfcond}).
Of course, in a real measurement the
 infinitesimal element of surface $dS$
is necessarily approximated by
a finite  surface $\Delta  S$, which may be for example the
sensor of some particle detector.
However, if one
wishes to measure the number of particles
traversing an arbitrary finite surface $S$, one should introduce the
concept of flux. This will be done in Subsection \ref{subsec:ddf}.

\subsubsection{Quantities related to the differential
  directional intensity}
Starting from the differential directional intensity $I_{ddi}$ it is
possible to construct several other quantities which are often
encountered in the literature. 

\begin{description}
\item[Integral directional intensity:] The integral directional
  intensity
(IDI)  $I_{idi}$ is
  obtained 
by integrating the  DDI over some finite
  interval of energy $\Delta E_i=E_{i,max}-E_{i,min}$:
\begin{equation}
I_{idi}(
\mathbf
r,
\theta,\phi, E_{i,min}, E_{i,max},t)
 =\int_{E_{i,min}}^{E_{i,max}} I_{ddi}(\mathbf
 r,\theta,\phi,E_i,t)dE_i
\label{idilab}
\end{equation}
where, of course, $E_{i,min}\ge 0$ in the non-relativistic case.
Moreover,
  $E_{i,max}\in[0,\infty]$.
The units of $I_{idi}$ are cm$^{-2}$s$^{-1}$sr$^{-1}$.
\item[Differential vertical intensity:]
In many experiments it is  measured the intensity of particles arriving on the
surface of the Earth from the vertical 
direction, i. e. the direction that from the ground
points toward the center of the earth~\footnote{Sometimes one  
  considers the intensity of particles 
  in the near vertical direction, where the values of
the angle $\theta$ between the trajectories of the
particles and the gravity force spans over a finite interval,
such as for instance  $0.9\le\sin\theta\le 1$.}. For this reason, the
DDI in the vertical direction, which is supposed here to
coincide with the value of the angle $\theta=0$ of our spherical
system of coordinates, has
deserved a separate name and it is called the  differential vertical
intensity (DVI) $I_{dvi}$.
The $I_{dvi}$ is defined as follows:
\begin{equation}
I_{dvi}(\mathbf
r,E_i, t) = I_{ddi}(\mathbf
r,
\theta=0,\phi,E_i,t) \label{idvi}
\end{equation}
The units of  DVI are the same of the
units of the DDI.
\item[Integral vertical intensity] The integral vertical intensity
  (IVI) gives the number of particles coming from the vertical
  direction with respect to our coordinate system $\varpi,\theta,\phi$
  and with energies comprised within the interval
  $[E_{i,min},E_{i,max}]$ which traverse a unit surface in the unit of
  time:
\begin{equation}
I_{ivi}(\mathbf r,t)=\int_{E_{i,min}}^{E_{i,max}}I_{ddi}(\mathbf
r,\theta=0,\phi,E_i,t) dE_i\label{ividef}
\end{equation}
\item[Integrated intensity:] The integrated intensity
  (II)\footnote{Strictly speaking the name integrated integral
  intensity would be more correct.} $I_{ii}$ is
  defined as 
  the integral of the DDI over all possible
  directions and energy values\cite{rossi}:
\begin{equation}
I_{ii}(\mathbf r,t)
=\int_0^{+\infty}dE_i\int_0^{2\pi}d\phi\int_0^{\pi}d\theta
\sin\theta 
 I_{ddi}
(
\mathbf
r,
\theta,\phi,E_i,t)
\end{equation}
The integrated intensity is measured in units cm$^{-2}$s$^{-1}$.
Of course, if one integrates the DDI only over all possible directions
the result is a quantity which may be called the
differential integrated intensity.
\end{description}

\subsection{Energy density of particles and the intensity in
  the relativistic case} 
The specific directional den\-si\-ty (SDD)
$u_{sdd}(\mathbf r,\theta,\phi,E_i,t)$  is defined here as
the kinetic energy of
particles of type $i$ per unit of volume, of energy and of solid
angle. 
Sometimes the SDD is also called differential directional density.
The quantity:
\begin{equation}
dU=u_{sdd}(\mathbf r,\theta,\phi,E_i,t) dVd\Omega dE_i\label{totkinene}
\end{equation}
represents the total kinetic energy  carried by particles which are
inside an element of volume $dV$ and have velocities $\mathbf v_i$
whose directions 
span a  small element of solid angle
$d\Omega$ centered around the direction of the unit vector $\mathbf
e_R(\theta, \phi)$. The norms of these velocities are determined by
the condition that 
the energy of the particles must be within the infinitesimal interval 
 $[E_i,E_i+dE_i]$. 
Clearly, the number of particles $dN$ with the above characteristics
which are
inside the small volume $dV$ at 
the time $t$ is given by the total energy of
the particles divided by the energy of each single
particle: 
\begin{equation}
dN=u_{sdd}(\mathbf r,\theta,\phi,E_i,t)dVd\Omega \frac{dE_i}{E_i}
\end{equation}
In the non-relativistic case we have seen that the relation between
the kinetic energy $E_i$ and the norm of the velocity $|\mathbf
v_i|$ is provided by Eq.~(\ref{kinene}). In the relativistic case,
this equation must be substituted with the following one:
\begin{equation}
|\mathbf v_i|=
\frac c{E_i+m_ic^2} \sqrt{E_i^2+2m_ic^2E_i}\label{relfor}
\end{equation}
The relativistic and non-relativistic expressions of the energy
density $u_{sdd}$ may be found in Ref.\cite{LandauLifsits}.

At this point, we wish to derive the DDI for
particles which attain relativistic speeds in terms of the energy density.
 We note to this purpose
that the particles which will traverse the surface $dS$ 
in the time $dt$ while arriving from the direction
perpendicular to $dS$, are contained in the infinitesimal volume:
\begin{equation}
dV=|\mathbf v_i|dSdt
\end{equation}
As a consequence, the number of particles $dN_i$ of Eq.~(\ref{defddi}) may be
expressed in terms of the SDD as follows:
\begin{equation}
dN_i=u_{sdd}(\mathbf r,\theta,\phi,E_i,t)|\mathbf v_i| dSdt
d\Omega
\frac{dE_i}{E_i} \label{dNiusdd}
\end{equation}
Comparing Eq.~(\ref{defddi}) with Eq.~(\ref{dNiusdd}), we obtain a
relation between the DDI and the SDD:
\begin{equation}
I_{ddi}=u_{sdd}(\mathbf r,\theta,\phi,E_i,t)
\frac c{E_i^2+m_ic^2E_i}\sqrt{E_i^2+2m_ic^2E_i}\label{biribiri}
\end{equation}
where we have used Eq.~(\ref{relfor}) in order to write the speed
$|\mathbf v_i|$ as a function of the energy $E_i$.
Eq.~(\ref{biribiri}) is the analogous of Eq.~(\ref{ddiexplfinal}) in
the relativistic case with the mass density $\rho_i$ replaced by the
energy density. 

At this point, the derivation of the related quantities of the DDI,
like for instance the integral directional intensity or the
differential vertical intensity, proceeds as in the non-relativistic
case of
Subsection~\ref{subsec:aaaa}. 

Finally, following an analogous calculation of the total energy
density of electromagnetic radiation presented in Ref.~\cite{ryblig},
it is possible to compute the total kinetic energy density
$u_{ted}(\mathbf r, t)$ of the particles per unit of volume:
\begin{equation}
u_{ted}(\mathbf r,t)=\int_{m_i^2c^4}^{+\infty}dE_i\int d\Omega
\frac
{E_i^2+m_ic^2E_i}
{c\sqrt{E_i^2+2m_ic^2E_i}
}
I_{ddi}(\mathbf r,\theta,\phi,E_i,t)
\end{equation}

\subsection{Differential directional flux and related
  quantities}\label{subsec:ddf} 
The definition of the differential directional flux (DDF) $\Phi_{ddf}$
is very
 similar to that of the DDI. The
 difference is that, in the case of the DDF,
 the
 direction of the normal $\mathbf n$ to the
element of surface  $d\mathbf S$ 
does not need to
 coincide with the direction of the velocity of the incoming particles
 $\mathbf e_R(\theta,\phi)$. 
 More precisely, the DDF is defined in such a way
that the quantity 
\begin{equation}
dN_{i,\mathbf e_R}^f=\Phi_{ddf}dSdtdE_id\Omega
\end{equation}
 represents the number of particles of a
given kind traversing
 the
infinitesimal surface element $dS$  
during the time $dt$ within the element of solid angle
  $d\Omega$ and within the energy interval $[E_i,
  E_i+dE_i]$. 
The superscript $f$ has been added to remember that now a flux is
being computed and not an intensity.

\begin{figure}[bpht]
\centering
\includegraphics[width=6cm]{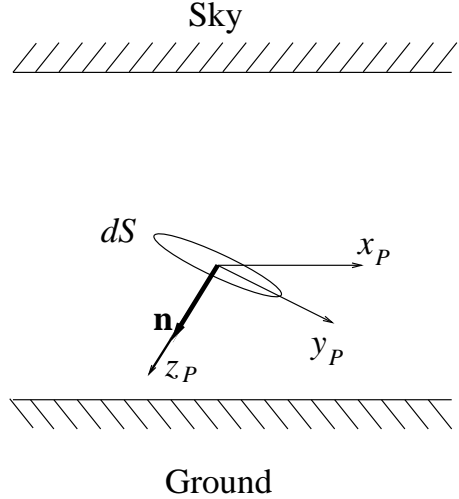}
\caption{This figure shows
  the schematic experimental setup used to measure the flux of
  particles of cosmic origin on the ground. The detector is
  represented as an infinitesimal element of surface $dS$. The active
  part of the detector is on the upper side of the surface, which
  points toward the sky.
}\label{schemedet}
\end{figure}

\begin{figure}[bpht]
\centering
\includegraphics[width=10cm]{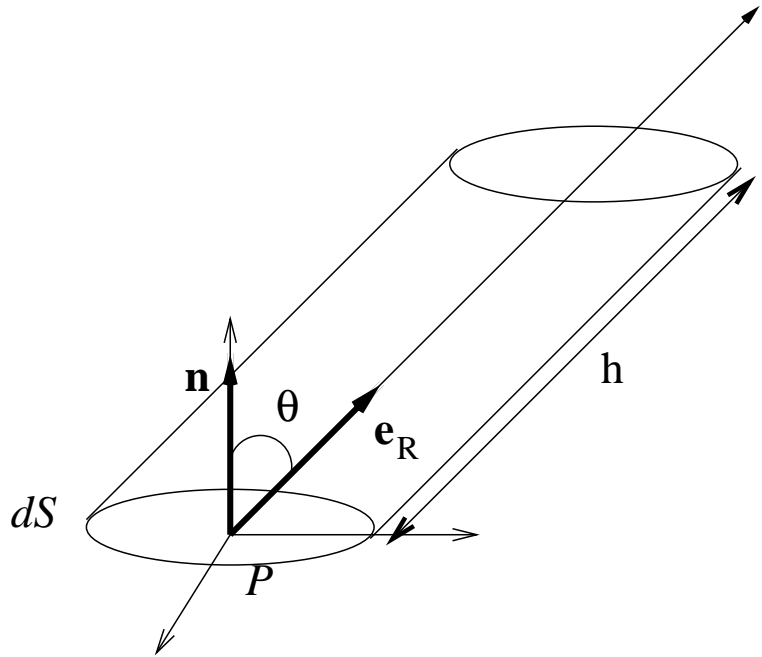}
\caption{This figure shows the geometrical setup for the definition of
  the differential directional flux. The normal vector  $\mathbf n$ to the
  infinitesimal surface $dS$ at the point $P$ makes an angle $\theta$
  with the
  vector $\mathbf e_R$ which gives the direction of the incoming particles.
The particles of type $i$ with velocity $|\mathbf v_i|\mathbf
  e_R(\theta,\phi)$ 
which will  traverse the surface $dS$ within the interval
  of time $dt$ are those contained in the volume $h\cdot
  dS\cos\theta$, where 
  $h=|\mathbf v_i|dt$.
}\label{flux}
\end{figure}

To compute the DDF, let us imagine that we wish to measure it in a
neighborhood of a point $P$. Since such measurements are usually
performed on the ground, to fix the ideas we assume that the point $P$
is very near (a few meters or less) to the surface of the Earth. The
particle detector is approximated as a small and flat surface, which is
centered around $P$. One side of the surface, that in which there are
the sensors and which is thus able to detect the fluxes of incoming
particles, is always directed toward the sky, while the opposite side
is pointing toward the ground, see Fig.~\ref{schemedet}.
To express this all in mathematical terms, we introduce an infinitesimal
element of surface $dS$ and we choose a system of coordinates
$x_P,y_P,z_P$ at the point $P$ in such a way that $dS$ lies in the
horizontal plane $z_P=0$. In polar coordinates $\varpi,\theta, \phi$
this means that the direction of the unit vector $\mathbf n$ which is
normal to $dS$ is given by the angle $\theta=0$. Furthermore, the
orientation of $\mathbf n$ is such that it points downward,
i. e. toward the ground. Since we are measuring only particles which
traverse the upper side of the surface $dS$ in a downward sense, this
implies that:
\begin{equation}
\mathbf n\cdot\mathbf e_R(\theta,\phi)=\cos\theta\ge 0\label{fluangthe}
\end{equation}
i.~e. the normal vector $\mathbf n$ and the
 particle velocity $|\mathbf v_i|\mathbf e_R(\theta,\phi)$ form an angle
 $\theta$, see Fig.~\ref{flux}.
Clearly, Eq.~\ceq{fluangthe} is satisfied only in the interval
 $0\le\theta\le\pi 2$
The volume of particles $dV$ which will traverse $dS$ coming from the
direction $\mathbf e_r(\theta,\phi)$ is shown in Fig.~\ref{flux} and
it is given by: $dV=|\mathbf v_i|dt\cos\theta$.
The number of those particle in the non-relativistic case is thus given by
\begin{equation}
dN_{i,\mathbf e_R}^f=\frac{\rho_i|\mathbf
  v_i|
}{m_i}dSdt\cos\theta
\end{equation}
 This is the analog of 
Eq.~(\ref{eeerer}) in the case of the flux.
 The factor $\cos\theta$ results after evaluating
the scalar product $\mathbf n\cdot \mathbf e_R(\theta,\phi)$ using
Eq.~(\ref{fluangthe}).
As in the case of the intensity, in real measurements
it is better to consider particles
coming from different directions and carrying different energies,
instead of
fixing the attention to a particular direction and a particular energy.
However we should remember that,
in the computation of the DDF, the orientation of the surface
$d\mathbf S$
remains fixed, what is changing is the direction of the incoming particles.
At this point the
procedure to obtain the differential directional flux is entirely
similar to that used in deriving the explicit
expression for the DDI of Eq.~(\ref{ddiexplfinal}). The final result
for the DDF is :
\begin{equation}
\Phi_{ddf}(\mathbf r,\theta,\phi,E_i,t)=
\frac{\rho_i}{(m_i)^\frac
  32}\sqrt{2E_i}\cos\theta \label{ddfexplfinal}
\end{equation}
The units of the DDF are the same as the units of the DDI.

\subsubsection{Observables related to the DDF}
In analogy with what has been done in the case of the intensities,
one may construct other observables starting from the DDF. For example,
after
integrating the DDF with respect to the energy one obtains the
{\bf integral directional flux} (IDF), while the {\bf differential vertical
flux} (DVF) 
corresponds to the value of the DDF in the case $\theta=0$. Clearly,
the DVF coincides with the DVI. 
The {\bf differential integrated flux }
$\Phi(\mathbf r,E_i,t)$,
which can be found in standard textbooks,
is defined in such a
way that the quantity $dN_{i,int}^f=
\Phi(\mathbf r,E_i,t)dSdt$ coincides with
the number of particles of a
given kind and of given energy $E_i$ traversing in a downward sense 
\footnote{We recall that downward means here that the angle between
the normal vector $\mathbf n $  
and the vector $\mathbf e_R$ which determines the direction of the
particle velocity must be within the range $[0,\frac\pi 2]$.}
 the
element of surface $dS$:
\begin{equation}
\Phi(\mathbf r,E_i,t)=
\int_0^{\frac{\pi}2}
 d\theta \int_0^{2\pi}d\phi \Phi_{ddf}
(
\mathbf r,
\theta,\phi,E_i,t)
\sin\theta
\end{equation}
The subscript $int$ in $dN_{i,int}^f$ means
integrated and it is referred to the fact that, to derive $\Phi(\mathbf
r,E_i,t)dSdt$, one needs an integration over $d\Omega$.
 This flux can be measured in units cm$^{-2}$s$^{-1}$GeV$^{-1}$.
Starting from the expression of $\Phi(\mathbf r,E_i,t)$ it is possible
to compute the integral integrated flux, often called {\bf total flux}:
\begin{equation}
\Phi(\mathbf r,t)=\int_{E_{i,min}}^{E_{i,max}}\phi(\mathbf r,E_i,t)dE_i
\end{equation}

The concept of flux is usually connected with vector fields.
In the present case, the vector field is provided by
the so-called  differential directional intensity field
\begin{equation}
\mathbf I_{ddi}=I_{ddi}\mathbf e_R
\end{equation}
where $\mathbf e_R$ is the unit vector which defines the direction of
the velocity of particles.
In terms of the DDI field, the flux may be expressed as follows:
\begin{equation}
\Phi(\mathbf r,E_i,t)=\int d\Omega \mathbf I_{ddi}\cdot
\mathbf n 
\end{equation}
according to the usual definition of flux.

Contrarily to the DDI,
 the DDF may also be
integrated with respect to the element of area $dS$. As we have
 anticipated before, this fact allows to
define the flux of particles traversing an
extended surface $S$. 
To compute the flux of particles in the case of an extended surface $S$,
it will be convenient to
parametrize this surface with the help of two parameters $\sigma_1$ and
$\sigma_2$, so that a point of $S$ in the space will be denoted by the
triplet of
cartesian coordinates 
$x(\sigma_1,\sigma_2)$, $y(\sigma_1,\sigma_2)$ and
  $z(\sigma_1,\sigma_2)$ or, shortly, by the radius vector
  $\mathbf r(\sigma_1,\sigma_2)$. 
For each point $P$ of the surface, corresponding to a given value of
the parameters $\sigma_1$ and $\sigma_2$, we have seen that
it is possible to compute
the differential flux 
of particles $\Phi(\mathbf r(\sigma_1,\sigma_2),E_i,t)$ traversing a
small element $dS$ of $S$. The total flux $\Phi_S(E_i,t)$ of particles
of energy $E_i$ incoming upon $S$ within the interval of time $dt$ is
obtained by 
integrating $\Phi(\mathbf r(\sigma_1,\sigma_2),E_i,t)$ with respect to
$dS$, where $dS$ will now depend on $\sigma_1$ and $\sigma_2$:
\begin{equation}
\Phi_S(E_i,t)=\int_S 
\Phi(\mathbf
r(\sigma_1,\sigma_2), E_i,t)dS(\sigma_1,\sigma_2) 
\label{totalflux}
\end{equation}

\end{appendix}


\begin{thebibliography}{99}
\bibitem{mewaldt} R. A. Mewaldt, {\it Cosmic Rays}, MacMillan
  Encyclopedia of Physics, J.~S.~Rigden Editor in Chief, MacMillan,
  New York 1996. 
\bibitem{pdg} S. Eidelman et al, {\it The Review of Particle Physics},
  {\it Phys. Lett.} {\bf B592} (2004), 1.
\bibitem{biermann} P. Biermann and G. Siegl, {\it Lect.Notes Phys.} {\bf
  576} (2001), 1. 
\bibitem{cronin} J. W. Cronin, {\it Rev. Mod. Phys.} {\bf 71} (1999),
  S165.
\bibitem{battistoni} G. Battistoni and A. F. Grillo,
{\it  Introduction to high energy cosmic ray physics}, 
published in the Proceedings of the
 ICTP School on Nonaccelerator Particle Astrophysics,
E. Bellotti, R.A. Carrigan, G. Giacomelli, N. Paver (Eds).
( Singapore, World Scientific 1996), 341.
Trieste, Italy, 17-28 Jul 1995. published
in Trieste 1995, Proceedings, Non-accelerator particle astrophysics,
341-374. 
\bibitem{Anchor} L. Anchordoqui et Al., {\it Int. Jour. Mod. Phys.}
  {\bf A18} (13) (2003), 2229, arXiv:hep-ph/0206072.
\bibitem{obrien} K. O'Brien et Al., {\it Env. Int.} {\bf 22} Suppl. 1
  (1996), S9.
\bibitem{nasaacr} Web page of the Laboratory for High Energy
  Astrophysics, http:/helios.gsfc.nasa.gov/acr.html.
\bibitem{gapnote} R. W. Clay, Z. Kurban and N. R. Wild, {\it Cosmic
  Ray Related Undergraduate Experiments}, GAP Note 1998-061 (Technical
  and Scientific  Notes about the Pierre Auger Project),
  http://www.auger.org/tech.html. 
\bibitem{RandKul} R. J. Rand and S. R. Kulkarni, {\it Ap. J.} {\bf
  343} (1989), 760.
\bibitem{lofar} J. P. Rachen and P. L. Bierman, {\it Astron. Astroph.}
  {\bf 272} (1993), 161.
\bibitem{exotic} P. Bhattacharjee and G. Sigl, {\it Phys. Rep.} {\bf
  327} (2000), 109.
\bibitem{cseven} J. W. Cronin, T. K. Gaisser and S. P. Swordy, {\it
  Sci. Am.} {\bf 276} (1) (1997), 32.
\bibitem{unscearb} UNSCEAR Report, {\it Sources and Effects of Ionizing
  Radiation, Annex B: Exposures from Natural Radiation Sources}
  (2000).
\bibitem{noelsta} Noel Stanton, {\it Introduction to cosmic rays},\\
http://www.phys.ksu.edu/{\~{}}evt/Quarknet/Docs/cosmic\_ray\_intro.pdf.
\bibitem{klecker} B. Klecker et Al., {\it Space Science Reviews} {\bf
  83} (1998), 259.
\bibitem{jokipii} J. R. Jokipii and J. Giacalone, {\it Space Science
  Reviews} {\bf 83} (1998), 123.
\bibitem{voy1} D. A. Zhuravlev et Al. {\it Cosmic Research} {\bf 43}
  (2005), 143.
\bibitem{voy2} N. A. Schwadron and D. J. McComas, {\it AIP Conference
  Proceedings} {\bf 719} (2004), 81.
\bibitem{voy3} S. M. Krimigis et Al., {\it Nature} {\bf 426} (2003), 45.
\bibitem{clay} R. W. Clay, A.G. K. Smith and J. L. Reid, {\it
  Publ. Astron. Soc. Aust.} {\bf 14} (1997), 195. 
\bibitem{milagro} C. Sinnis et Al. {\it Teravolt Astrophysics -- The
  Milagro Gamma-Ray Observatory}, 2003 Physics Division Activity
  Report, 145,
Los Alamos National Laboratory,\\
http://www.lanl.gov/p/prog\_rpt.shtml
\bibitem{doereport} U. S. Department of Energy (DOE), 
{\it Radiological control technician training}, DOE Handbook
DOE-HDBK-1122-99, Module 1.07,
Washington 1999. This manuscript is available on-line at the URL:\\
http://www.eh.doe.gov/techstds/standard/hdbk1122/rad.html
\bibitem{allan} H. R. Allan, {\it Prog. in Elem. Part. and Cos. Ray
  Phys,}
J. G. Wilson and S. A. Wouthuysen (Eds.) N. Holland Publ. Co, Vol {\bf
  10} (1971), 171.
\bibitem{lofara} H. Falcke and P. Gorham, {\it Detecting Radio Emission
  from Cosmic Ray Air Showers and Neutrinos with LOFAR}, LOFAR
  Scientific Memorandum \#3, 29 May 2002.
\bibitem{eheas} X. Bertou, M. Boratav and A. Letessier-Selvon, {\it
  Physics of extremely energetic cosmic rays}, arXiv:astro-ph/0001516.
\bibitem{TalkPierog} T. Pierog, R. Engel and D. Heck, {\it Impact of
  Uncertainties in Hadron Production in Air Shower Predictions},
  transparencies of a talk presented at the Conference {\it From
  Colliders to Cosmic Rays}, Prague (CZ), 7--13 September 2005:\\
http://www.particle.cz/conferences/c2cr2005/prog.html.
\bibitem{allkofer} O. C. Allkofer et Al., {\it Phys. Lett.} {\bf 36B}
  (1971), 425; O. C. Allkofer, {\it Introduction to Cosmic Radiation},
(Verlag Karl Thiemig, M\"unchen, Germany, 1975).
\bibitem{dar} A. Dar, {\it Phys. Rev. Lett.} {\bf 51} (1983), 227.
\bibitem{schumacher} R. A. Schumaker, {\it Cosmic Ray Muons}, material
  of the course ``Modern Physics Laboratory'' held at the Carnegie
  Mellon University:\\
http:/www-meg.phys.cmu.edu/physics\_33340
\bibitem{desalin} D. Desilets and M. Zreda, {\it Earth and
  Plan. Science Lett.} {\bf 193} (2001), 213.
\bibitem{ziegler} J. F. Ziegler, {\it Nucl. Inst. Meth.} {\bf 191}
  (1981), 419.
\bibitem{usat}
U.S. Committee on Extension to the Standard Atmosphere (COESA),
{\it U.S. Standard Atmosphere, 1976}, published by the 
U.S. Government Printing Office, Washington, D.C.
\bibitem{zieglerb} J. F. Ziegler, {\it IBM Jour. Res. Develop} {\bf
  42} (1) (1998), 117.
\bibitem{fowler} J. W. Fowler et Al., {\it Astropar. Phys.} {\bf 15}
  (2001), 49.
\bibitem{kitpmini} S. Koutu, {\it The Pierre Auger Observatory:
  Progress and Perspectives}, talk delivered at the 
KITP Miniprogram on  Astrophysics of Ultra-High Energy Cosmic Rays,
  Photons, and Neutrinos,\\
http://online.kitp.ucsb.edu/online/uhe05/
\bibitem{ssareport} SSP 30512 revision C,
{\it Space Station Ionizing Radiation Design Environment},
International Space Station Alpha Program (1994).
\bibitem{ryblig} G. B. Rybicki and A. P. Lightman, {\it Radiative
  Processes in Astrophysics}, John Wiley and Sons, (New York,
  Chichester, Brisbane, Toronto, Singapore 1979).
\bibitem{rossi} B. Rossi, {\it Rev. Mod. Phys.} {\bf 20} (1948), 537.
\bibitem{LandauLifsits} L. D. Landau and E. M. Lifshitz, {\it The
  Classical Theory of Fields}, (Pergamon Press, 1975).
\end{thebibliography}
\end{document}